\documentclass[journal=jacsat,manuscript=article]{achemso}

\usepackage[version=3]{mhchem} 
\usepackage{afterpage}
\usepackage{textgreek}

\author{Taran Driver}
\author{Vitali Averbukh}
\author{Leszek J. Frasi\'nski}
\author{Jon P. Marangos}
\author{Marina Edelson-Averbukh}
\affiliation{The Blackett Laboratory, Imperial College London, London SW7 2AZ, UK}
\email{m.edelson-averbukh@imperial.ac.uk}

\title{Two-dimensional partial covariance mass spectrometry for the top-down analysis of intact proteins}

\begin{document}



\begin{abstract}
 Two-dimensional partial covariance mass spectrometry (2D-PC-MS) exploits the inherent fluctuations of fragment ion abundances across a series of tandem mass spectra, to identify correlated pairs of fragment ions produced along the same fragmentation pathway of the same parent (e.g. peptide) ion. Here, we apply 2D-PC-MS to the analysis of intact protein ions in a standard linear ion trap mass analyzer, using the fact that the fragment-fragment correlation signals are much more specific to bio-molecular sequence than 1D MS/MS signals at the same mass accuracy and resolution. We show that from the distribution of signals on a 2D-PC-MS map it is possible to extract the charge state of both parent and fragment ions without resolving the isotopic envelope. Furthermore, the 2D map of fragment-fragment correlations naturally reveals the secondary decomposition pathways of the fragment ions. We access this spectral information using an adapted version of the Hough transform. We demonstrate the successful identification of highly charged, intact protein molecules without the need for high mass resolution. Using this technique we also perform the \textit{in silico} deconvolution of the overlapping fragment ion signals from two co-isolated and co-fragmented intact protein molecules, demonstrating a viable new method for the concurrent mass spectrometric identification of a mixture of intact protein ions from the same fragment ion spectrum.
\end{abstract}


In a tandem mass spectrometry (MS/MS) experiment, an ensemble of the biomolecules under analysis (e.g. peptides \cite{Medzihradszky2005} or oligonucleotides \cite{Schurch2007}) is introduced into the gas phase in ionic form, isolated, and fragmented. The abundances and mass-to-charge ($m/z$) ratios of the resultant fragments are measured, and this information is used to piece together the original structure of the analyzed molecule. Two-dimensional partial covariance mass spectrometry \cite{Driver2020Two-DimensionalCorrelations,PhysToday} (2D-PC-MS) extracts an additional dimension of information from the same experimental observable: by correlating the inherent, random fluctuations in the scan-to-scan abundances of different fragment ions the technique is able to identify pairs of fragment ions which are produced in the same or consecuitive decomposition pathways of the same parent molecule. This additional information has been shown \cite{Driver2020Two-DimensionalCorrelations} to dramatically improve the specificity of fragment signal matching for database searches and to correctly identify biologically important mixtures of combinatorially modified histone isomers \cite{Driver2020histones}. In this work, we extend of 2D-PC-MS to intact proteins within top-down mass spectrometry. 

Top-down mass spectrometry \cite{Toby2016} is a rapidly growing field which is primarily driven by a desire to maximize the proteoform \cite{Smith2013a} coverage. Here we develop the top-down capability within 2D-PC-MS by using an adapted form of the Hough transform, a well-known computer vision algorithm \cite{Duda1972} orignally formulated to enable the automatic analysis of bubble chamber pictures \cite{Hough1959}. Applying the Hough transform to the two-dimensional partial covariance maps of protein decomposition allows us to detect series of 2D-PC-MS features related to the same parent ion. This opens the possibility to use 2D-PC-MS for so-called `multiplex' measurements \cite{Masselon2000,Chapman2014} of intact protein molecules, in which multiple parent ions are fragmented and measured concurrently. We demonstrate multiplex 2D-PC-MS analysis through \textit{in silico} deconvolution of the strongly overlapping fragment ions from the co-fragmentation of two different, highly charged protein ions.

`Top-down mass spectrometry' refers to the tandem mass spectrometric analysis of intact protein ions, bypassing the enzymatic digestion step in the canonical bottom-up proteomic workflow. Important holistic information on the sequence and its post-translational modifications (PTMs) is often destroyed at the enzymatic digest step in bottom-up workflows \cite{Zhang2013}. Following sample preparation, the first step in top-down mass spectrometry is the introduction of the molecules under analysis into the gas phase in ionized form, e.g. by electrospray ionisation (ESI) \cite{Fenn1989,Toby2016}. Due to the larger number of residues available to be protonated/deprotonated on longer sequences the gaseous ions of intact protein molecules typically appear at higher charge states than peptide ions. This significantly complicates the analysis of both the molecular mass of the protein itself and its fragment ion spectra.

The fragmentation spectra of intact protein ions contain a complex mix of ions at different charge states, requiring isotopically resolved fragment measurements for their correct interpretation. The canonical method for the identification of fragment ion charge states in 1D MS/MS is to exploit the small natural abundance of heavier isotopes (e.g. \textsuperscript{13}C at $\sim 1.1\%$ and \textsuperscript{15}N at $\sim 0.4\%$) to determine the charge of a fragment ion, by measuring the $m/z$ difference between different isotopic peaks of the same molecule. Consecutive peaks are separated in $m/z$ by $\frac{1}{z}$, where $z$ is the charge of the fragment ion. In order to correctly infer the charge state of the fragment ion from the isotopic envelope, an accurate and well-resolved measurement of the $m/z$ difference between peaks in the isotopic envelope is required. The necessary resolving power is available from Fourier transform-based mass analyzers such as Fourier transform ion cyclotron resonance (FT-ICR) and Orbitrap analyzers \cite{Scigelova2011}, or quadrupole time-of-flight mass analyzers \cite{Morris1996}. The linear ion trap, desirable for its speed, sensitivity and affordability \cite{Douglas2005}, has been used in top-down proteomics \cite{Bunger2008,Kim2005,Diedrich2011}, but its significantly lower mass resolution and accuracy limit its applicability within the top-down paradigm. Practically. the relatively low mass resolution of the linear ion trap mass analyzer means, for example, that fragment ions of charge 5+ or greater are typically impossible to identify \cite{Kim2005}, because the isotopic envelope is unresolved. 

Furthermore, even when measured at high mass resolution, the complex overlapping fragment peaks resulting from the multitude of different fragmentation pathways available to highly charged intact protein sequences can be extremely difficult to interpret. One of the most challenging consequences of directly fragmenting an intact protein molecule is the inevitable production of large numbers of so-called `internal fragments', containing neither terminus of parent protein molecule, regardless of the fragmentation method used \cite{Durbin2015,Savaryn2016,Li2018}. These internal ions, typically resulting from the secondary fragmentation of larger terminal ions, greatly increase the difficulty of protein identification from top-down fragment ion spectra \cite{Breuker2008,Cannon2014,Durbin2015,Xiao2017}. The vast number of possible internal products from one protein structure means their use for structural analysis is strongly limited, but they are ubiquitous in experimental spectra and so constantly risk being incorrectly interpreted as different fragment ions of the wrong sequence.

Two-dimensional partial covariance mass spectrometry (2D-PC-MS) is based on calculating the self-correcting partial covariance map of the MS/MS spectrum across multiple repeated scans \cite{Driver2020Two-DimensionalCorrelations}. By identifying signals in the MS/MS spectrum which synchronously rise and fall in intensity across repeated measurements, 2D-PC-MS is able determine pairs of fragment ions which were produced along the same or consecuitive fragmentation pathways of the same molecule. These signals can be identified by positive islands of partial covariance on the 2D-PC-MS map. The partial covariance represents the correlation between two signals once spurious `common-mode correlations' -- which cause all spectral signals to correlate to all others as a result of global scan-to-scan fluctuations in external experimental parameters -- have been suppressed \cite{Frasinski2016}. Within 2D-PC-MS, this is done by using the `total ion count' (TIC) as a single partial covariance parameter derived from the spectrum itself, representing a viable proxy to the compound effect of the many experimental parameters which fluctuate from scan to scan.

The fidelity of correlation islands appearing on a 2D-PC-MS map is assessed using the 2D-PC-MS correlation score, which is calculated by normalising the volume of a 2D-PC-MS island to the standard deviation of that volume upon jackknife resampling. This metric enables the identification of even extremely low volume true correlation features from higher volume statistical noise \cite{Driver2020Two-DimensionalCorrelations}. Correlation signals from a 2D-PC-MS map can be ranked according to the value of their 2D-PC-MS correlation score, and it is often instructive to plot the high-ranking true correlation features as a scatter plot \cite{Driver2020Two-DimensionalCorrelations}.

A significant feature of 2D-PC-MS is that it provides the direct experimental identification of sequence-specific complementary ion pairs. These are pairs of fragment ions formed by cleavage of a single bond in the same parent ion, such as the b/y pairs commonly produced in the collisional-induced activation (CID) of peptide and protein ions \cite{Wysocki2005}, and are typically the favoured fragment ions for sequence identification. 
In 2D-PC-MS, correlations between different sets of complementary ion pairs, formed from the cleavage of different bonds in the same parent ion, arrange themselves along so-called `mass conservation lines' \cite{Driver2020Two-DimensionalCorrelations}.  Due to the conservation of mass and charge, all complementary ion pairs resulting from cleavage of a different bond in the same parent ion lie along lines defined on the 2D-PC-MS map by:
\begin{equation}
y = -\frac{z_1}{z_2} \times x + \frac{M_{P}}{z_2} \,,
\label{eqn:mcline}
\end{equation}
where $z_1$ is the charge state of the fragment ion correlated on the first ($x$-) axis, $z_2$ is the charge state of the fragment ion correlated on the second ($y$-) axis, and $M_{P}$ is the mass of the parent ion. Eq.~(\ref{eqn:mcline}) describes a straight line with gradient $-\frac{z_1}{z_2}$ and y-intercept $\frac{M_{P}}{z_2}$. Hence, if it is possible to determine a set of 2D-PC-MS correlation islands lying on a particular mass conservation line, then both the ratio between the fragment charge states $\frac{z_1}{z_2}$ and the value of $\frac{M_{P}}{z_2}$ can be identified. Provided even limited knowledge on the charge state of the parent ion (e.g. an upper limit), this enables calculation of the charge state of each of the correlated fragment ions, as well as the charge state and mass of the parent ion that gave rise to the mass conservation line. 

Complementary ion pairs are particularly useful in protein sequence identification because they are reliably sequence-specific \cite{Kryuchkov2013}. MS/MS measurements are at risk of false positive sequence identifications due to measured peaks from chemical or electrical noise, or non-canonical fragment ions produced by cleavages of the parent ion which are typically unaccounted for. The most prevalent non-canonical ions across nearly all fragmentation techniques are internal fragment ions \cite{Wysocki2005,Bunger2008,Brodbelt2016}. Whilst internal fragments can be extremely useful for e.g. the identification and localisation of PTMs \cite{Driver2020Two-DimensionalCorrelations}, they are particularly problematic in the analysis of intact proteins \cite{Breuker2008,Cannon2014,Durbin2015,Xiao2017}. This is partly because the probability for the incorrect assignment of internal ions explodes for longer sequances, since the number of different internal ions possible for a given parent ion grows as the square of the parent sequence length. This renders identification of complementary fragment ions, whose number grows only linearly, particularly valuable for top-down analysis.

Given prior knowledge of the parent ion mass and charge, it is possible to readily predict the set of mass conservation lines for complementary ions originating from fragmentation of a particular parent molecule using Eq.~(\ref{eqn:mcline}). However, it is also common to observe the mass conservation lines along which the mass of the two correlated ions does not sum to the mass of the intact molecule. Examples of such cases are the secondary fragmentation of a terminal fragment ion into a smaller terminal fragment ion and internal fragment ion, or the correlation between two terminal fragment ions whose mass sum to the mass of the parent ion, minus a small neutral loss (e.g. H\textsubscript{2}O, NH\textsubscript{3} or CO). For such mass conservation lines, an \textit{a priori} definition of their positions using the high-resolution measurement of the intact protein ion is not practical, but their correct identification would provide valuable structural and mechanistic information. Therefore, we have developed a numerical technique to locate mass conservation lines on the 2D-PC-MS maps without an \textit{a priori} knowledge of the mass and charge state of the parent ion. To this end, we use a version of the Hough transform adapted to identify the mass conservation lines defined by different sets of ion pair correlations on a 2D-PC-MS map. We term this adapted version, whose full implementation is described in the Materials and Methods, the `restricted Hough transform'. Once the signals along the mass conservation lines are identified, one can use them as an input for the 2D-PC-MS automatic database search engine \cite{Driver2020search_engine} in order to define the protein sequence. 

\section*{Materials and Methods}
\subsection*{2D-PC-MS}
The technique of 2D-PC-MS is described in detail in reference \cite{Driver2020Two-DimensionalCorrelations}. Within 2D-PC-MS the measured fragment mass spectrum at every scan is treated as a row vector $\mathbf{X} = [X_1, X_2, \ldots, X_n]$, with $\mathbf{Y} = [Y_1, Y_2, \ldots, Y_n]$ being a column vector fragment intensities. Each element of the vectors $\mathbf{X}$ and $\mathbf{Y}$ varies from scan-to-scan, and the self-correcting partial covariance is given by:
\begin{equation}
pCov(\mathbf{Y}, \mathbf{X}; TIC) = Cov(\mathbf{Y}, \mathbf{X}) - \frac{Cov(\mathbf{Y},TIC)Cov(TIC,\mathbf{X})}{var(TIC)};
\label{eqn:pcovtic}
\end{equation}
\begin{equation}
Cov(\mathbf{Y}, \mathbf{X})= \langle \mathbf{YX} \rangle- \langle \mathbf{Y} \rangle \langle \mathbf{X} \rangle,
\label{eqn:cov}
\end{equation}
where $TIC$ represents the total integrated ion count at each scan (deduced from the spectrum itself) and $var(TIC)$ is the variance of this total ion count across the full set of scans. Subtraction of the second term in Eq. \ref{eqn:pcovtic} removes the uninteresting intensity correlations induced by the global scan-to-scan fluctuations inherent in MS/MS measurements.

Equation \ref{eqn:pcovtic} renders a partial covariance matrix:
\begin{center}
$\begin{bmatrix}
    pCov(Y_1,X_1) & pCov(Y_1,X_2) & \dots  & pCov(Y_1,X_l) \\
    pCov(Y_2,X_1) & pCov(Y_2,X_2) & \dots  & pCov(Y_2,X_l)  \\
    \vdots & \vdots & \ddots & \vdots \\
    pCov(Y_m,X_1) & pCov(Y_m,X_2) & \dots  & pCov(Y_n,X_n)
\end{bmatrix}$,
\end{center}
where the element at $(i, j)$ represents the TIC partial covariance between element $i$ of vector $\mathbf{Y}$ and element $j$ of vector $\mathbf{X}$. This is typically visualised as a partial covariance map, and positive islands on this map identify two fragments which were born in the same or consecuitive fragmentations of the same biomolecule. Automated analysis of the 2D-PC-MS maps developed in reference \cite{Driver2020Two-DimensionalCorrelations} returns pairs of correlated $m/z$ values, analagous to the single $m/z$ values returned by standard 1D MS/MS analysis. Unlike the relative fragment abundance by which the fidelity of 1D MS/MS signals is measured, the fidelity of 2D-PC-MS signals is assessed by the 2D-PC-MS correlation score $S(x,y)$ \cite{Driver2020Two-DimensionalCorrelations}, which has been shown to provide a robust measure of whether a signal on the 2D-PC-MS map represents a true correlation even for very low intensity self-correcting partical covariance signals distinguishing them from the statistical noise, e.g. due to finite summations in eq.~\ref{eqn:pcovtic}. This score is calculated according to:
\begin{equation}
S(x,y)=\frac{V(x,y)}{\sigma[V(x,y)]},
\label{eqn:corrscore}
\end{equation}
where $V(x,y)$ is the volume of a peak on the partial covariance map between $m/z$ values $x$ and $y$ and $\sigma[V(x,y)]$ is the standard deviation of this volume across a series of resamples of the original scan set (this is performed by jackknife resampling) \cite{Driver2020Two-DimensionalCorrelations}. Higher 2D-PC-MS correlation scores identify higher fidelity 2D-PC-MS signals. In Figs.~\ref{fig:myoglobin} and \ref{fig:topdownmix1.1}, the top-ranking 2D-PC-MS signals according to S(x,y) have been plotted on a scatter plot.

\subsection*{Hough Transform}
The Hough transform provides a computationally efficient algorithm for feature detection in image analysis \cite{Hough1959,Duda1972}, and although since generalised to identify the positions of arbitrary shapes \cite{Ballard1981} it was originally developed to identify lines, which is how it is used here. Specifically, we use it to find straight lines passing through multiple 2D-PC-MS correlation islands. In this implementation, it works by constructing an accumulator array in the Hesse normal $\rho, \theta$ parameter space for straight lines, and identifying maxima in this accumulator space as lines which pass through multiple points.

\begin{figure}[tbhp]
\centering
\includegraphics[width=87mm]
{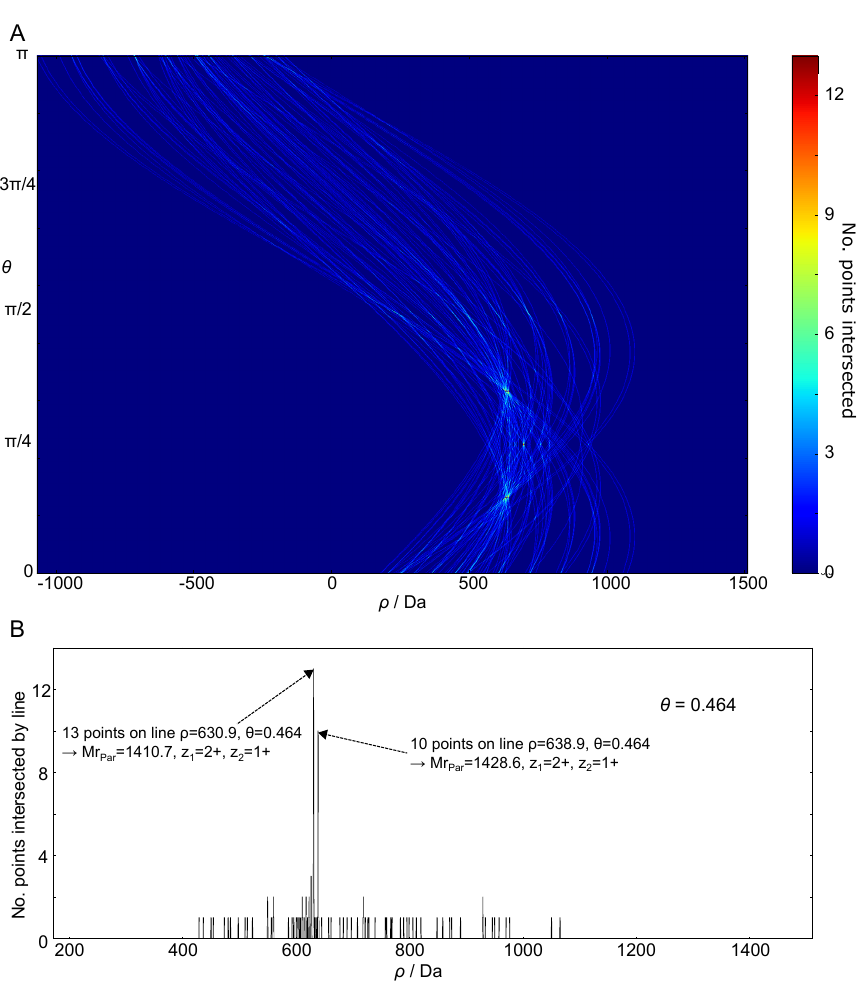}
\caption{\small Accumulator spaces for the Hough transform. \textbf{a}, 2D accumulator space, plotted as a colour map, constructed for all points on the  2D-PC-MS map in Fig.~\ref{fig:mclines_cs}. The path in $\rho, \theta$ parameter space that is traced out by the set of lines passing through each point takes a sinusoidal form. The value of the accumulator space at each $\rho, \theta$ is the number of points through which the corresponding straight line passes. \textbf{b}, 1D accumulator space plotted for $\theta=0.464$. Because the mass conservation lines can only appear at particular gradients, the Hough transform to identify mass conservation lines was performed by scanning the one-dimensional parameter space spanned by $\rho$ at each possible value for $\theta$. Restricting $\theta$ also restricts the possible values that $\rho$ can take, as can be seen on the x-axis of this plot compared to the x-axis of panel \textbf{a}.}
\label{fig:houghtrans}
\end{figure}

In general, for each point in an image (e.g. a 2D-PC-MS map), the parameter space spanned by the accumulator is scanned and for every $\rho, \theta$ describing a line closer than a given distance from the point in question, the corresponding element of the accumulator space is incremented by one. Each point traces out a sinusoidal curve in the full 2D $\rho, \theta$ accumulator space, as shown in panel \textbf{a} of Fig.~\ref{fig:houghtrans}, in which the colour map of the accumulator for all points shown on the 2D-PC-MS map in Fig.~\ref{fig:mclines_cs} is plotted. This process is repeated for each point on the map, so that at the end of the procedure the value of each element of the accumulator is equal to the number of points through which the straight line described by the corresponding $\rho, \theta$ values passes. Therefore, all lines passing through a certain number of points or more can be directly read off the accumulator.

For the purposes of identifying mass conservation lines on a 2D-PC-MS map, it is possible to greatly reduce the computational work required to perform the Hough transform. Since all mass conservation lines have gradient $-\frac{z_1}{z_2}$, where $z_1$ and $z_2$ each correspond to a discrete charge state and one may choose that $z_1 \geq z_2$, mass conservation lines are restricted to a limited set of gradients and therefore a limited set of corresponding $\theta$ values. As such, only a very small subsection of the full parameter space need be scanned to identify all mass conservation lines, namely only those slices in $\theta$ which correspond to a possible gradient of a mass conservation line. Restriction of $\theta$ also allows for a tightening of the limits for $\rho$. This restricted Hough transform was implemented as set of consecutive scans across the one-dimensional $\rho$ space, each at a different set value of $\theta$ (see panel \textbf{b}, Fig. \ref{fig:houghtrans}), and was found to significantly increase the computational speed of the procedure ($\sim240 \times$ speed-up for the 2D-PC-MS map in Fig.~\ref{fig:mclines_cs}, $\Delta \theta=1.0^\circ$ for the non-restricted Hough transform). Due to the symmetry of 2D-PC-MS maps, every map features two correlation islands corresponding to the same physical correlation. For the case of mass conservation lines where $z_1=z_2$, both such correlation islands fall on the same mass conservation line because $-\frac{z_1}{z_2} = -\frac{z_2}{z_1} = -1$. Therefore, for $\theta=\frac{\pi}{4}$, the values of the accumulator are halved prior to further processing in order that they represent the true number of physical correlations lying along a mass conservation line. Thanks to the Hesse normal parametrization, the restricted Hough transform is also able to identify the vertical or horizontal series of correlation islands that appear when one fragment ion correlates with another fragment ion and a series of derivative ions due to neutral losses (see e.g. Fig.~2 in reference \cite{Driver2020Two-DimensionalCorrelations}). 

\subsection*{MS Analysis}
All samples were prepared to a concentration of $\sim$1 \textmu M in a solution of 1\% formic acid, 50\% acetonitrile and 49\% water. All solvents were of Optima\textsuperscript{TM} LC/MS grade and purchased from Fisher Scientific Ltd. Myoglobin from equine skeletal muscle, cytochrome c from equine heart and ubiquitin from bovine erythrocytes were all purchased from Sigma-Aldrich. All experiments were performed on a Thermo Fisher Scientific LTQ XL linear ion trap mass spectrometer. The samples were directly infused using a a Harvard Apparatus 11 Plus Single Syringe Pump coupled to a Nanospray II Ion Source (Thermo Fisher Scientific) at 1 \textmu l/min. The electrospray source voltage was manually optimised on spray stability at each measurement between 1.8 and 2.2 kV, with no auxiliary desolvation gas applied. The temperature of the ion transfer capillary was held constant at 200$^\circ$C. The parent ions under analysis were isolated at an AGC MS\textsuperscript{n} target value of 300 (to augment the scan-to-scan fluctuations and maximise the signal-to-noise on the 2D-PC-MS map), and fragmented at a normalised collision energy of 70\%. 
Fragment ion measurements were acquired in the linear ion trap at a scan rate of 125,000 Da/sec.

\subsection*{Data Analysis}
The `.raw' files produced by the LTQ XL were converted to text files by the Thermo Xcalibur File Converter software. All subsequent analysis was performed by software written in-house using the Python language. The input text file is read into the numerical `.npy' format, and used to calculated the partial covariance map according to eq.~\ref{eqn:pcovtic}. The software identified the highest 3000 features on the 2D-PC-MS map, and calculated the 2D-PC-MS correlation score $S(x,y)$ for each of these features according to eq.~\ref{eqn:corrscore}. It produced as output a list of $m/z$ pairs correlated on the 2D-PC-MS map along with the corresponding 2D-PC-MS correlation score of the correlation island. See reference \cite{Driver2020Two-DimensionalCorrelations} for more details.

The identification of complementary pairs is performed by a program which accepts as input the list of 2D-PC-MS correlation islands with 2D-PC-MS correlation scores, the measured $m/z$ value of the isolated and fragmented parent ion(s) and, the range of expected charge states for the measured ions. It expects the parent $m/z$(s) to be measured on a linear ion trap mass analyzer, resulting in high uncertainty in $m/z$ and no identification of the charge state. The restricted Hough transform described above is then applied to the `image' described by the top $N$ 2D-PC-MS score-ordered correlation islands to identify all straight lines, with admissible gradient for a mass conservation line, along which more than a thrsehold value of correlations fall. In order to remove mass conservation lines due to the further fragmentation of a primary dissociation product (such as a terminal ion dissociating to an internal ion and smaller terminal ion), which are useful but not desired for the purposes described here, the parent ion mass and charge described by all such mass conservation lines is then queried to determined whether it would produce an $m/z$ value within a tolerance (here, 1.5 Da) of the measured parent $m/z$. If it does, the mass conservation line is determined to be a primary mass conservation line along which complementary ions of the intact parent ion are located. Finally, a simple clustering algorithm ($\Delta m/z = 3$ Da) is applied to the primary mass conservation lines to determine all lines from the same parent ion. The program outputs individual files with all primary correlations from the same parent ion, along with the charge state of the fragment ions and charge state and mass of parent ion, determined by the software. For all measurements in this work, points within 1.5 Da of a line were determined to lie along it and a threshold of 6 points was applied to the accumulator space for determination of a mass conservation line. 

\section*{Results and Discussion}

\begin{figure}[tbp]
\centering
\includegraphics[width=\textwidth]
{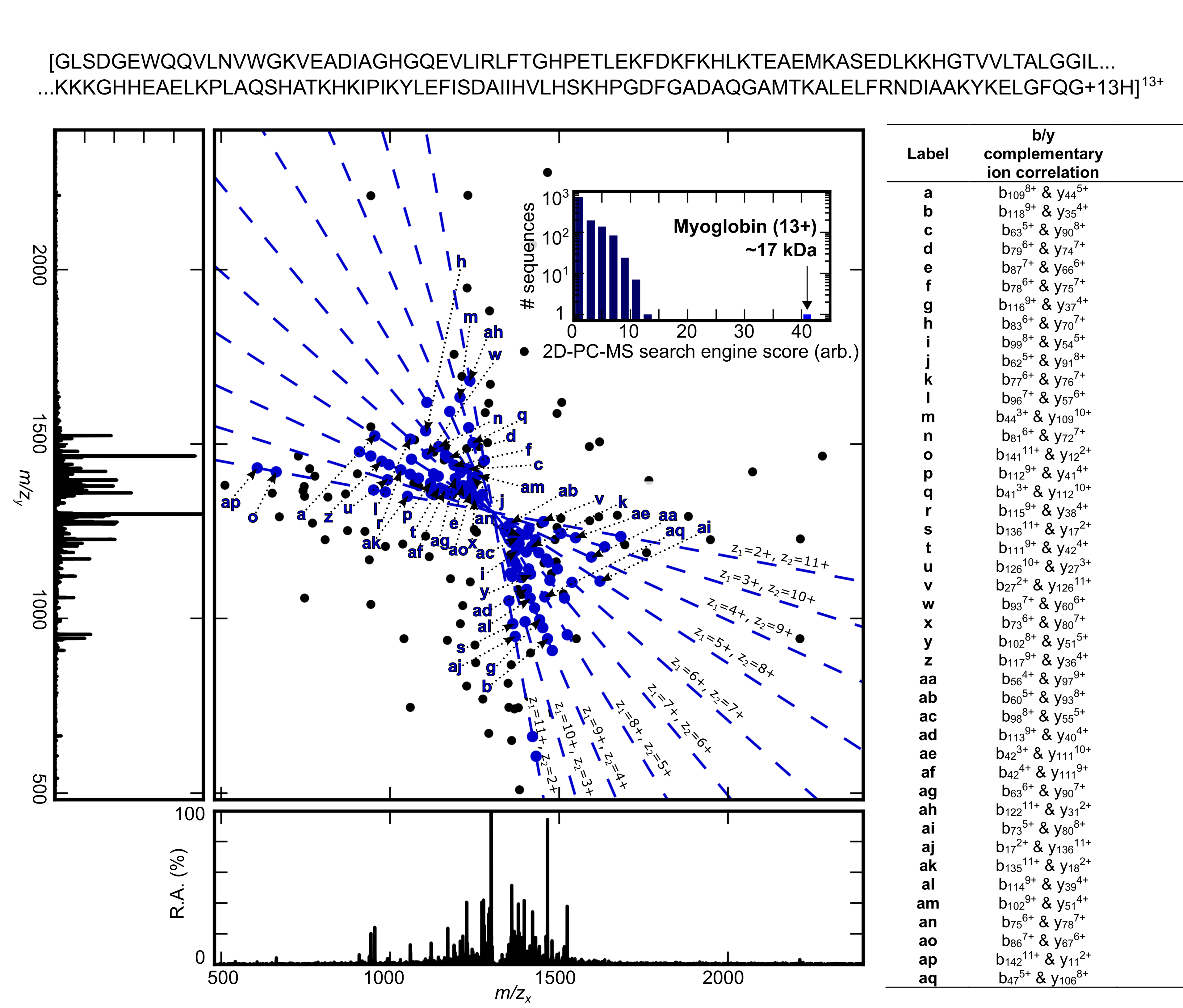}
\caption{\small 2D-PC-MS map of 13+ ion of intact protein myoglobin from equine skeletal muscle ($\sim$17 kDa, sequence shown above map). The top 100 2D-PC-MS correlation score-ordered (see Materials and Methods) signals are plotted on the scatter plot. The blue dashed lines are the mass conservation lines identified by the restricted Hough transform, showing different charge partitions across the complementary ion correlations which fall along them (blue dots). Identities of the annotated complementary ion correlation signals, which are each labelled on only one side of the autocorrelation line for visual clarity, are given in the table to the right. The alphabetical order of the annotated correlations reflects their ranking according to the 2D-PC-MS correlation score. The 2D-PC-MS search engine has correctly identified the intact protein ion from its top-down 2D measurement by assigning it a dominant similarity score (see inset histogram), using the charge state and $m/z$ value of the the intact protein ion and its complementary fragments measured directly from the map. Note the logarithmic scale on the y-axis of the histogram.}
\label{fig:myoglobin}
\end{figure}

Figure \ref{fig:myoglobin} shows the 2D-PC-MS measurement of the intact protein myoglobin from equine skeletal muscle (molecular weight $\sim$17 kDa, 153 amino acid residues long). The 2D-PC-MS correlation signals, obtained in LTQ-XL  measurement of myoglobin (13+) are presented as a scatter plot. These features were passed as an input to the restricted Hough transform (see Materials and Methods) which has identified 5 symmetrically inequivalent mass conservation lines. Due to the symmetry of the maps each mass conservation line appears twice, with respective duplicates mirrored in the line $m/z_x=m/z_y$. The 5 inequivalent mass conservation lines correspond to 5 different charge partitions (e.g., 9+/4+, 8+/5+, etc.)  for the complementary ions from a parent ion of mass ($M_p$ in eq.~\ref{eqn:mcline}) 16958.0 Da and charge 13+. This parent ion mass, determined directly from the application of the restricted Hough transform to the 2D-PC-MS map, shows a deviation of 0.04\% from the expected theoretical mass. 
The the fragment-fragment correlations lying on the identified mass conservation lines have been fed into the 2D-PC-MS search engine, which which compares theoretical fragment-fragment correlations of candidate sequences to the measured (experimental) fragment correlation signals.
The inset of Fig.~\ref{fig:myoglobin} displays the results of the candidate sequence matching producing the outstandingly highest score for the correct protein sequence. 
All correlations falling along the identified mass conservation lines are plotted in blue, and the 43 annotated signals have all been identified by the 2D-PC-MS database search engine
matching procedure as complementary b$_a^{z_1+}$ \& y$_b^{z_2+}$ ion pairs where $a+b=153$ and $z_1+z_2=13$. 2D-PC-MS matches fragment ions from top-down measurements according to their calculated and predicted isotope-averaged, rather than monoisotopic masses, as described in the Supporting Information. Because the restricted Hough transform requires a series of correlation signals belonging to the same mass conservation line to identify that particular line, its performance has been found to improve with growing sequence length. This is because the number of possible correlation signals belonging to a particular mass conservation line grows linearly with sequence length. 

\subsection*{\textit{In Silico} Deconvolution of Mixtures of Intact Proteins by 2D-PC-MS} 

\begin{figure}[tbhp]
\centering
\includegraphics[width=0.85\textwidth]
{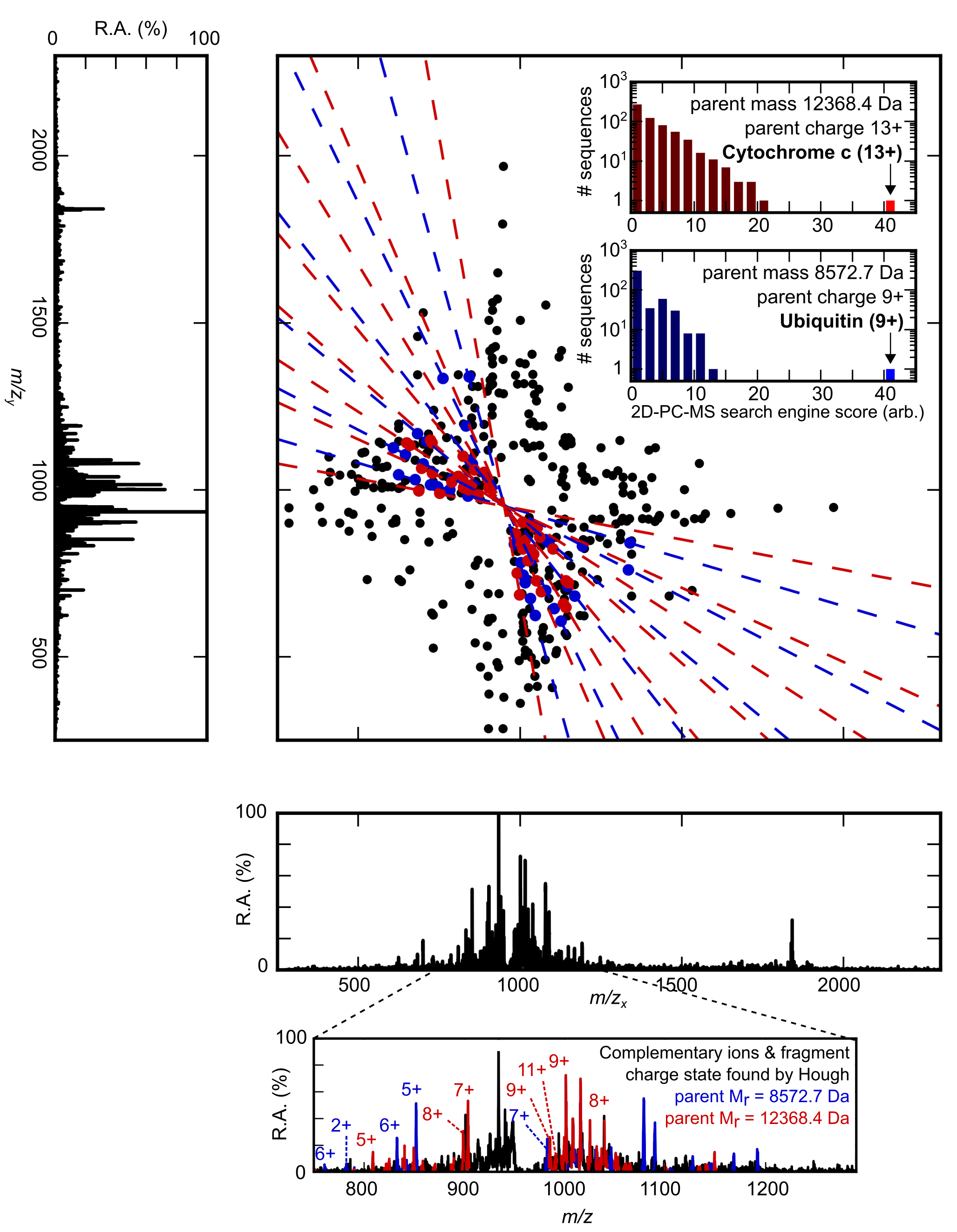}
\caption{\small \textit{In silico} deconvolution of the fragment spectrum of co-isolated and co-fragmented protein parent ions cytochrome c (13+, $m/z=951.4$) and ubiquitin (9+, $m/z=952.5$) using 2D-PC-MS. The $x,y$ positions of the plotted fragment ion correlations are provided to the restricted Hough transform to identify mass conservation lines in the correlation map. From the congested overlapping fragment ion spectrum of the parent ion mixture, the restricted Hough transform has discovered two different sets of mass conservation lines (red and blue) along which lie the complementary fragment ion correlations of two different intact protein ions. Both protein ions have been identified by the 2D-PC-MS search engine using the separated fragment ion correlations, parent masses and parent charge states, all of which have been determined by the restricted Hough transform. Note the logarithmic scale on the y-axis of the histograms.}
\label{fig:topdownmix1.1}
\end{figure}

Biological samples are typically complex mixtures of more than one protein, and separation of these mixtures prior to top down 1D MS/MS analysis is essential to avoid the insurmountably difficult task of identifying proteins from the overlapping 1D fragment ion signals resulting from the simultaneous decomposition of several protein molecules. 
Multiple protein separation techniques have been developed but they nevertheless frequently result in protein co-fragmentation  \cite{Catherman2014}. 
2D-PC-MS allows for the \textit{in silico} separation of protein mixtures which have been co-isolated and co-fragmented, without the costly, wasteful and challenging process of upstream separation. Co-isolation and co-fragmentation refers to the concurrent isolation, fragmentation, and fragment ion measurement for different molecular structures in the same analysis step, which usually produces fragment ion spectra of intractable complexity. According to Eqn. \ref{eqn:mcline}, the complementary ions produced by the fragmentation of parent molecules of different mass and/or charge state fall along uniquely defined mass conservation lines. The separation of overlapping fragment ions direct from the 2D-PC-MS map therefore requires the identification of the different mass conservation lines present. As demonstrated in Fig. \ref{fig:myoglobin}, this is straightforwardly performed by the restricted Hough transform. 

Figure \ref{fig:topdownmix1.1} demonstrates the \textit{in silico} separation of the two co-isolated and co-fragmented intact protein ions, cytochrome c (13+) and ubiquitin (9+). Plotted are the top 200 2D-PC-MS correlation score-ranked features, which have been passed to the restricted Hough transform along with the roughly determined parent ion $m/z$ values as measured in the precursor scan in the linear ion trap. The restricted Hough transform has identified two sets of mass conservation lines, corresponding to parent ions of average mass 8572.7 Da\footnote{Deviation of 0.01\% from theoretical mass of the ion.} and charge state 9+ (blue) and average mass 12368.4 Da\footnote{Deviation of 0.04\% from theoretical mass of the ion.} and charge state 13+ (red). The zoomed-in view of the horizontal 1D MS/MS spectrum illustrates the deconvolution and charge state identification performance of the restricted Hough transform in one region of the spectrum. Each set of correlation features lying along the two different sets of mass conservation lines has been individually passed to the 2D-PC-MS search engine, along with the parent mass and charge state as determined by the restricted Hough transform. As illustrated by the inset histograms in Fig.~\ref{fig:topdownmix1.1}, the 2D-PC-MS search engine unambiguously identifies each of the two mixed proteins from the two sets of deconvolved 2D-PC-MS features. Both sets of mass conservation lines in Fig.~\ref{fig:topdownmix1.1} cross the $m/z_x=m/z_y$ diagonal at almost the same point. This is a result of the fact that both co-fragmented parent ions have very similar $m/z$ values (951.5 and 952.5), chosen such in order to to achieve protein co-fragmentation in the linear ion trap.

\section*{Conclusions}
In conclusion, we have successfully applied the new technique of 2D-PC-MS to the tandem MS measurement of intact protein molecules. To perform the top-down analysis an adapted version of the Hough transform was used to identify the complementary fragment ion correlation in the 2D-PC-MS maps. Interestingly, this information allows one to determine the charge state of these fragment ions without the requirement of resolving the isotopic envelopes of the signals. We were able to identify intact proteins from top-down measurements made on a linear ion trap mass spectrometer despite its modest mass resolution. We have also used 2D-PC-MS to successfully deconvolve, in silico, the complementary fragment ions from the highly complex chimera spectrum of two different protein ions, allowing for high-confidence identification of both co-isolated proteins. Because 2D-PC-MS is a general technique depending only on instrumental detection efficiency the methodology described here should be immediately applicable to the top-down analysis, including chimera spectra deconvolution on other mass spectral platforms. 

\newpage




\begin{suppinfo}

\paragraph{2D-PC-MS Determination of Fragment and Parent Ion Charge State and Parent Ion Mass} 

The phenomenon of `mass conservation lines' is a direct result of the conservation of mass and charge. Take a correlation between a complementary ion pair at $m/z$ values $m_{1}/z_{1}$ \& $m_{2}/z_{2}$. Due to the fact the same MS/MS spectrum appears on each axis of the 2D-PC-MS map, this same physical correlation produces exactly two correlation islands on the 2D-PC-MS map -- at $x=m_{1}/z_{1}, y=m_{2}/z_{2}$ and $x=m_{2}/z_{2}, y=m_{1}/z_{1}$ -- but for the following discussion we consider the set of correlation islands  described by $x=m_{1}/z_{1}, y=m_{2}/z_{2}$, where for each physical correlation $z_{1}$ and $z_{2}$ are determined by $z_{1} \geq z_{2}$. This set covers all physical correlations measured on the 2D-PC-MS map.
If the the ions measured at $m_{1}/z_{1}$ \& $m_{2}/z_{2}$ are a complementary ion pair from a parent ion with mass $M_{P}$, then the sum of the masses of the complementary ions will be equal to the mass of their parent ion. This means that the following equality holds true:
\begin{equation}
(m_{1}/z_{1}) \times z_1 + (m_{2}/z_{2}) \times z_2 = M_{P}
\end{equation} 
Therefore, all correlation islands which correlate the primary products from the decomposition of a parent ion of mass $M_{P}$ will fall along one of the straight lines of the form:
\begin{equation*}
(x \times z_1) + (y \times z_2) = M_{P}
\end{equation*}
which can be rearranged to:
\begin{equation}
y = -\frac{z_1}{z_2} \times x + \frac{M_{P}}{z_2},
\label{eqn:mcline2}
\end{equation}
and therefore defines a straight line with gradient $-\frac{z_1}{z_2}$ and y-intercept $\frac{M_{P}}{z_2}$. 

\begin{figure}[tbhp]
\centering
\includegraphics[width=87mm]
{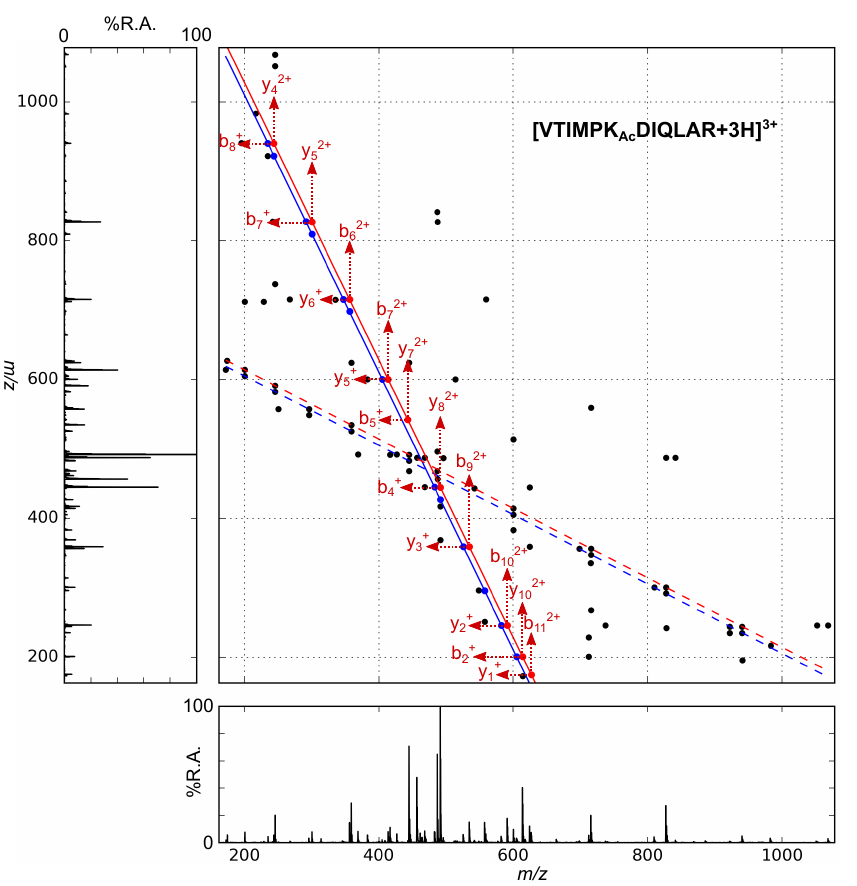}
\caption{\small Scatter plot of the top 50 2D-PC-MS correlation score-ordered correlation islands from the 2D-PC-MS map of the dissociation of the triply protonated ion of peptide VTIMPK\textsubscript{Ac}DIQLAR, showing the mass conservation lines formed due to the dissociation of the parent molecule (red), and dissociation of the parent molecule minus small neutral loss (blue). The averaged 1D mass spectra are plotted along the each side of the 2D-PC-MS scatter plot.}
\label{fig:mclines_cs}
\end{figure}

Thus, all complementary ion pairs from a dissociating parent ion\footnote{As well as the intact precursor ion this could be e.g. a terminal fragment ion which further fragments to a series of terminal ion/internal ion correlations.} of mass $M_{P}$ and charge $Z_P=z_1+z_2$ will appear on the 2D-PC-MS map along a straight line of gradient $-\frac{z_1}{z_2}$ and y-intercept $\frac{M_{P}}{z_2}$. For the reciprocal set of correlation islands where $x=m_{2}/z_{2}, y=m_{1}/z_{1}$ and $z_{2} \geq z_{1}$, all correlation islands will fall on a straight line with gradient $-\frac{z_2}{z_1}$ and y-intercept $\frac{M_{P}}{z_1}$. A useful corollary of the above is that if it is possible to determine a set of correlation islands lying on a particular mass conservation line, both the ratio between the charge states $\frac{z_1}{z_2}$ and the value of $\frac{M_{P}}{z_y}$ can be identified. Provided limited knowledge on the charge state of the parent ion (e.g. an upper limit), this generally enables the charge state of each of the correlated fragment ions, as well as the charge state and mass of the parent ion responsible for the mass conservation line (if not already known), to be inferred.

Fig. \ref{fig:mclines_cs} shows a scatter plot of the top 50 2D-PC-MS correlation score-ordered correlation islands from the 2D-PC-MS map of triply protonated peptide ([VTIMPK\textsubscript{Ac}DIQLAR+3H]\textsuperscript{3+}) under CID at normalised collision energy 35\%. There are 10 correlation islands (red points) lying along the red line, which has gradient=-2 and y-intercept=1428.6. Following the above, all correlation islands lying along this line with coordinates $x=m_{1}/z_{1}, y=m_{2}/z_{2}$ result from the correlation of two ions with $m_1/z_1$ and $m_2/z_2$, where $z_1=2+$ and $z_2=1+$. Additionally, from the gradient and $y$-intercept of this mass conservation line all 10 correlation islands are inferred to have arisen from the dissociation of a parent molecule of charge $Z_P=z_1+z_2=3+$ and mass $M_{P}= z_1 \times $y$-intercept= 1 \times 1428.6 =1428.6$ Da, which corresponds to the mass of the M+3H parent ion (expected mass 1428.8 Da). This was confirmed by the manual assignment of all 10 correlation islands along the mass conservation line, with fragment ion charge state verified using the isotopic envelope in the 1D mass spectrum. The blue line, along which 13 blue correlation islands are plotted, has gradient=-2 and $y$-intercept=1410.7. This corresponds to a parent ion of charge 3+ and $M_P=1410.7$, i.e. the correlation of two primary products from the parent ion minus neutral loss of a water molecule (expected at $1428.8 - 18.0 = 1410.8$). The assignments of these correlation islands are not plotted for visual clarity. The two dashed lines show the mass conservation lines for the reciprocal set of correlation islands where $x=m_{2}/z_{2}$ and $y=m_{1}/
z_{1}$.

\paragraph{Average masses}
Within 2D-PC-MS, the repeat MS/MS scans are performed at a high scan rate in order to increase the number of scans it is possible to take in a given amount of time, providing better statistics for calculation of the partial covariance. The decrease in mass accuracy and resolution associated with this high scan rate is more than compensated for by spectral decongestion along the second dimension, the improved specificity of 2D correlations \textit{vs.} 1D fragment ion signals and the ability to determine fragment ion charge states without resolution of the isotopic envelope. For peptides and oligonucleotides, we have developed a robust method for determination of the monoisotopic $m/z$ values of the two correlated fragments ions, which provides an accuracy well within the standard $\pm$ 0.8 Da $m/z$ tolerance for standard linear ion trap measurements \cite{Driver2020Two-DimensionalCorrelations}. However, Fig. \ref{fig:topdownaccscat} demonstrates the systematic inaccuracy accumulated by these determined monoisotopic masses for fragment ions at higher $m/z$. For the fragment ions of correlations which are able to be automatically assigned by the 2D-PC-MS search engine software (and so fall within the specified $m/z$ tolerance of $\pm$0.8 Da), the fragment ion $m/z$ is plotted on the $x$-axis whilst the deviation of that value from the expected (theoretical) $m/z$ for that value is plotted on the $y$-axis. \begin{figure}[tbhp]
\centering
\includegraphics[width=87mm]
{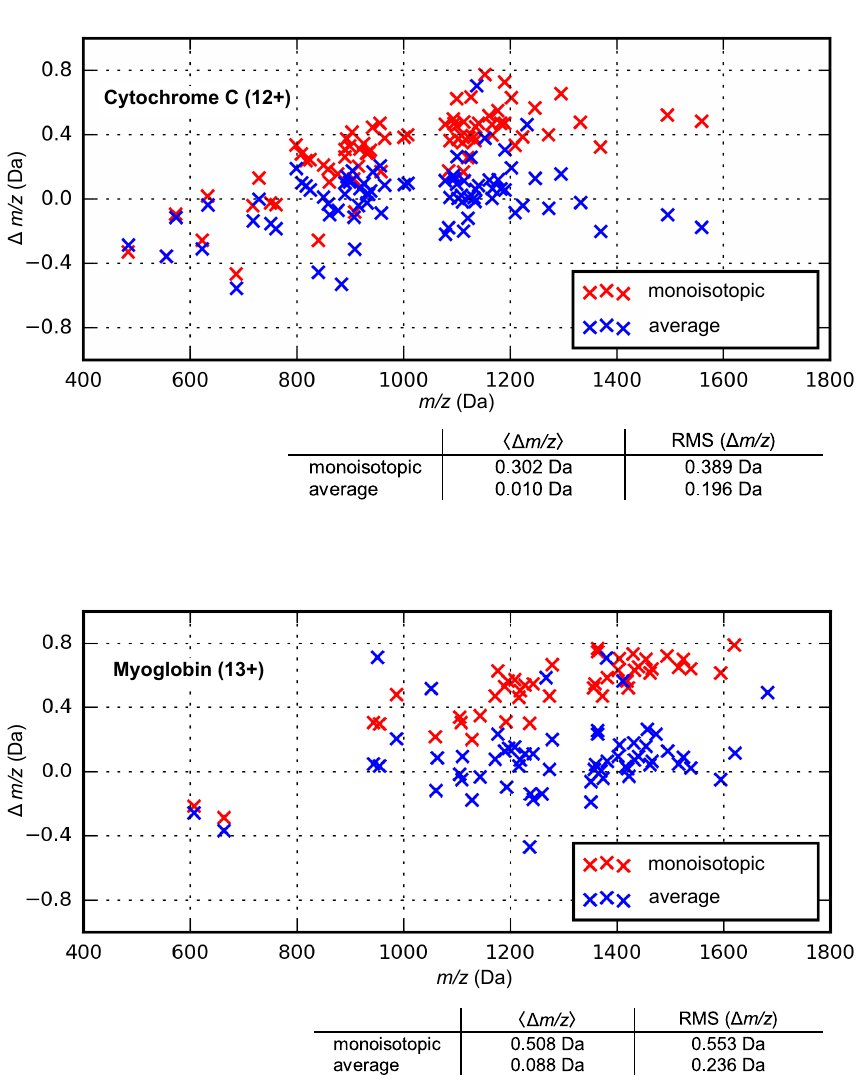}
\caption{\small Deviations in $m/z$ for matched 2D-PC-MS correlations in top down measurements, monoisotopic mass (red) \textit{vs.} average mass (blue).}
\label{fig:topdownaccscat}
\end{figure} The determined monoisotopic $m/z$'s show a steadily increasing deviation with increasing $m/z$ for both the parent ions plotted. This shift is expected given the method of determination of monoisotopic mass employed; which uses the centre-of-mass of the isotopic distribution as a reference. The centre of mass of the isotopic distribution shifts up and further away from the monoisotopic mass as the mass of the measured ion increases, due to the greater relative abundance of heavier isotopic species. However, by taking the centre-of-mass of all 2D-PC-MS peaks and matching these values with the $m/z$ values corresponding to theoretical average masses instead of the theoretical monoisotopic masses, it is possible to achieve consistently high $m/z$ accuracy which remains robust across the entirety of the mass spectrum (blue crosses is Fig. \ref{fig:topdownaccscat}). The $m/z$ values for all correlated fragment ions in this work have been identified in this manner, leading to high confidence identification of all measured protein ions.

\paragraph{Search engine}
The principles of operation of the 2D-PC-MS search engine will be detailed in an upcoming publication \cite{Driver2020search_engine}.
It works by matching the measured 2D-PC-MS spectrum to theoretical 2D-PC-MS spectra generated \textit{in silico} from a database of possible sequences. The experimentally measured 2D-PC-MS correlation signals are weighted according to their 2D-PC-MS correlation score. If a database sequence produces an outstanding similarity score with the measured spectrum, it is deemed to be the correct sequence. For the top-down 2D-PC-MS search engine database searches performed in this work, the specified parent mass tolerance was 0.1\%. The searches were performed against the UniProtKB/Swiss-Prot database \cite{Bateman2017} with no enzymatic digest (i.e. only intact protein sequences), with N-terminal acetylation, heme C and initiator methionine loss all specified as variable modifications. The search was performed using the measured and theoretical average fragment ion $m/z$ values, at a fragment ion matching tolerance of $\pm$0.8 Da.

\end{suppinfo}

\bibliography{references}

\end{document}